\documentclass[twoside,aps,superscriptaddress,pre,showpacs,twocolumn,floatfix,citesort,a4page]{revtex4}
\usepackage{graphicx}
\usepackage{dcolumn}
\usepackage{amssymb,amsmath}
\usepackage{epsfig}
\usepackage[dvips]{color}
\usepackage{lscape}
\usepackage{longtable}
\usepackage{rotating}

\newcommand\euro{{\sffamily C%
    \makebox[0pt][l]{\kern-.70em\mbox{--}}%
    \makebox[0pt][l]{\kern-.68em\raisebox{.25ex}{--}}}~}

\usepackage{graphicx}
\usepackage{bm}
\usepackage{amssymb}
\usepackage{color}
\usepackage{amsmath}

\newcommand{\be}{\begin{equation}}
\newcommand{\ee}{\end{equation}}

\begin{document}
\title{Velocity gradients statistics along particle trajectories in turbulent flows:\\ 
the refined similarity hypothesis in the Lagrangian frame}

\author{Roberto Benzi} \affiliation{Dept. Physics and INFN, University
  of Tor Vergata, Via della Ricerca Scientifica 1, 00133 Rome, Italy.}

\author{Luca Biferale} \affiliation{Dept. Physics and INFN, University
  of Tor Vergata, Via della Ricerca Scientifica 1, 00133 Rome, Italy.}

\author{Enrico Calzavarini} 
  \affiliation{Lab.\ de Physique, \'Ecole Normale Sup\'erieure de Lyon,
CNRS UMR5672, 46 All\'ee d'Italie, 69007 Lyon, France.}

\author{Detlef Lohse} \affiliation{Dept. of Science and Technology, Impact Institute, and Burgers Center,
  University of Twente, 7500 AE Enschede, The Netherlands.}

\author{Federico Toschi} \affiliation{Dept. Physics and Dept. Mathematics \& Computer Science, 
   Eindhoven University of Technology, P.O. Box 513, 5600 MB Eindhoven, The Netherlands.}

\collaboration{International Collaboration for Turbulence Research}
\date{today}

\begin{abstract}
  We present an investigation of the statistics of velocity
  gradient related quantities, in particluar energy dissipation rate and enstrophy, along the trajectories of fluid tracers and of
  heavy/light particles advected by a homogeneous and isotropic
  turbulent flow. The Refined
  Similarity Hypothesis (RSH) proposed by  Kolmogorov and Oboukhov in 1962 is rephrased in the Lagrangian context
  and then tested  along
  the particle trajectories. The study is performed on state-of-the-art
  numerical data resulting from numerical simulations up to  $Re_\lambda \sim
  400$ with $2048^3$ collocation points. When particles have small inertia, we show
  that the Lagrangian formulation of the RSH is well verified
  for time lags larger than the typical response time $\tau_p$ of the particle.
  In contrast, in the large inertia limit when the particle response time approaches the integral-time-scale of the flow,   particles behave
   nearly ballistic, and the Eulerian formulation of RSH holds  in the inertial-range.
\end{abstract}
\date{\today}

\maketitle

\section{Introduction}
One of the most prominent features of turbulent flows is the strong
variation present in the energy dissipation field, a phenomenon called
intermittency \cite{frisch:1995}. In an attempt to describe
quantitatively intermittent fluctuations in the inertial range of
turbulence, Kolmogorov and Oboukhov in 1962
\cite{kolmogorov:1962,oboukhov:1962} proposed a general relation
linking velocity fluctuations, measured at a given spatial increment
$\delta_r u = u( x+r,t)-u(x,t) $, with the statistical properties of
the coarse grained energy dissipation, $\varepsilon_r = r^{-3}
\int_{\Lambda(r)} {\varepsilon}({\bm x},t) \;d^3x $ averaged over a
volume, $\Lambda(r)$, of typical linear size $r$:
 \begin{equation}
   \delta_r u \sim r^{1/3} \varepsilon_r^{1/3},
   \label{eq:erksh}
\end{equation}
where $\sim$ means ``scales as" or ``equal in law''.
Equation (\ref{eq:erksh}) is known as the Refined (Kolmogorov)
Similarity Hypothesis (RSH) and it is considered to be one of the most
remarkable relations between turbulent velocity fluctuations: Many efforts  in the last decades have been devoted
to its validation
(\cite{stolovitzky:1994,chen:1997,toschi:2000}).  The
importance of RSH cannot be underestimated: it bridges inertial-range
properties with small-scale properties, supporting the existence of an
energy cascade mechanism, statistically local in space. So far, a
rather strong evidence supports the validity of the RSH in the
Eulerian frame (i.e. the laboratory frame). On the other hand, no
investigation has been reported in the literature on the validity of
RSH in the Lagrangian frame (i.e. along fluid particle trajectories). 
The main difficulty in studying RSH in a moving reference frame stems from the necessity to make
  multi-point measurements along particle trajectories in order to
  calculate  the stress tensor. As a result, no
  experimental measurements along particle trajectories of velocity
  gradients exists for time long enough to be able to evaluate temporal
  correlations. Also numerical experiments are very demanding,
  requiring refined computations of velocity differences along particle
  trajectories. This is usually implemented by computing the velocity gradients matrix in Fourier space,  then transforming it to physical space by  (inverse) Fast Fourier Transform, and performing off-grid interpolations of the gradients at the particle positions.
  Here we report the first of such measurements using high-resolution Direct Numerical Simulations (DNS) investigations.
We also note that when the particles transported in a turbulent environment have non-negligible size or mass, i.e. they are inertial paricles, their
trajectories becomes strongly sensitive to the statistical and topological properties of the advecting flow (\cite{eaton:1994,calzavarini:2008a,calzavarini:2008b,toschi:2009}). The possible validity of Lagrangian RSH in this context is far from being trivial and may
shed new light on the physics of particulate transport in turbulent flows: an ubiquitous phenomena in nature and in industrial
applications alike.

In the present study we will extend the RSH relation to the temporal domain and test its validity along the trajectories of fluid
tracers and of inertial particles whose density is smaller/larger than the fluid one while their sizes span the interval from the dissipative to the inertial range of scales.
The manuscript is organized as follows: First we gives details on the numerical methods of the DNS. We then present the extension of RSH to the Lagrangian domain and we test it on the trajectories of fluid tracers. In the last section we investigate the case of inertial particles: we show under which conditions the Lagrangian RSH still holds and how it should be modified in the special case of highly-inertial particles.

\section{Numerical methods}
The incompressible fluid velocity $\bm u(\bm x, t)$, $\bm \nabla \cdot \bm u = 0$, evolves according to
the  Navier-Stokes equations :
\begin{equation} {{{\rm D}\bm u} \over
    {{\rm D} t}} \equiv {{\partial \bm u} \over {\partial  t}} + \bm u \cdot \bm
  \nabla \bm u = -  \frac{\bm \nabla p}{\rho_f} +\nu \Delta \bm u + \bm f,\label{eq:ns}
\end{equation} 
where $p$ denotes the pressure, $\rho_f$ the fluid density assumed constant, and $\bm f$ an external large-scale forcing
injecting energy at a mean rate $\langle \varepsilon \rangle =\langle \bm u\cdot \bm
f\rangle$. 
Together with the Eulerian field we integrated the Lagrangian evolution of fluid tracers: ${{{\rm d}\bm x(t)} / {{\rm d} t}}=\bm v \equiv \bm u(\bm x(t), t)$,
and point-particles by means of a model of dilute suspensions of small
passively advected spherical particles, as derived in
Refs.~\cite{maxey:1983,gatignol:1983,auton:1988}:
\begin{equation} 
{{{\rm d}\bm x} \over {{\rm d} t}}=\bm v\,,\qquad
  {{{\rm d}\bm v} \over {{\rm d} t}} =\beta \, {{{\rm D}\bm u} \over
    {{\rm D} t}} + {1\over {\tau_p}}(\bm u -\bm v)\,,
\label{eq:dynamics}
\end{equation}
where $\bm x$ and $\bm v$ denote the particle position and velocity,
respectively.  
In Eq. (\ref{eq:dynamics}) the coefficient $\beta=3\rho_f/(\rho_f+2\rho_p)$
is related to the ratio between the density of the particle ($\rho_p$) and of the
fluid ($\rho_f$); $\tau_p=a^2/(3\beta\nu)$ is
the particle response time, with $a$ the particle-radius. 
The Stokes number of the particle is defined as $St=\tau_p/\tau_{\eta}$, where $\tau_\eta \equiv ( \nu / \langle \varepsilon \rangle )^{-1/2} $ is the dissipative time-scale of the turbulent flow.
In our simulation the parameters $\beta$ and $St$ can be varied independently, therefore it is possible to consider also the case $(\beta=0, St >  0)$, corresponding to the limit to very heavy particles for which the fluid added mass is negligible while Stokes drag is the only relevant dynamical force. On the other hand the situation $(\beta=1, St = 0)$ is equivalent to the case of a perfect fluid tracer.

Eq.~(\ref{eq:ns}) is numerically integrated by means of a standard internally 2/3 de-aliased pseudo-spectral algorithm with a second order Adams-Bashforth time-advancing scheme. The very same time-scheme is used to track the particles evolving according to eq. (\ref{eq:dynamics}): the time-step size in both cases is $O(10^{-2} \tau_{\eta})$, however particle informations are recorded for post-processing/analysis at a rate of $10^{-1} \tau_{\eta}$.
Interpolations of the velocity field, acceleration field (necessary for (\ref{eq:dynamics})) and velocity gradients at the particle positions, are done via a tri-linear algorithm.  For a validation of our numerical method we address Ref. \cite{calzavarini:2009}, where a satisfactory comparison on acceleration lagrangian statistics has been performed against an independent numerical implementation with several different features (field interpolation based on tri-cubic scheme, external dealiasing procedure, a slightly different large scale forcing). 
In our DNS energy is injected at large-scale by maintaining the spectral content of the first two
shells in Fourier space constant.  Here we will report data coming from two
sets of simulations with $N^3=2048^3$ and $N^3=512^3$ collocation
points, corresponding to $\mathrm{Re}_\lambda=400$ and $180$,
respectively, and sampling the parameter space $\beta\in[0:3]$, $St \in
[0:4]$ with $64$ ($\beta,St$) particles types. 
A total amount of $\sim 10^{8}$ particles are tracked in time. 
Results on the clustering of these
particles in the turbulence have already been reported in Ref.\ \cite{calzavarini:2008a,calzavarini:2008b}.
Inertia requires some time before particles reach their fractal (or multifractal) statistically stationary distributions \cite{bec:2003,bec:2005}. We
therefore waited till the Lagrangian statistics became stationary (approximately one large-eddy turnover-time) before performing the analysis presented here.  Measurements of velocity differences and gradients are based on sets of $O(10^6-10^7)$ particles which have been followed in time for few $O(1)$ large-eddy-turnover times.  
 
\begin{figure*}
  \includegraphics[width=\hsize]{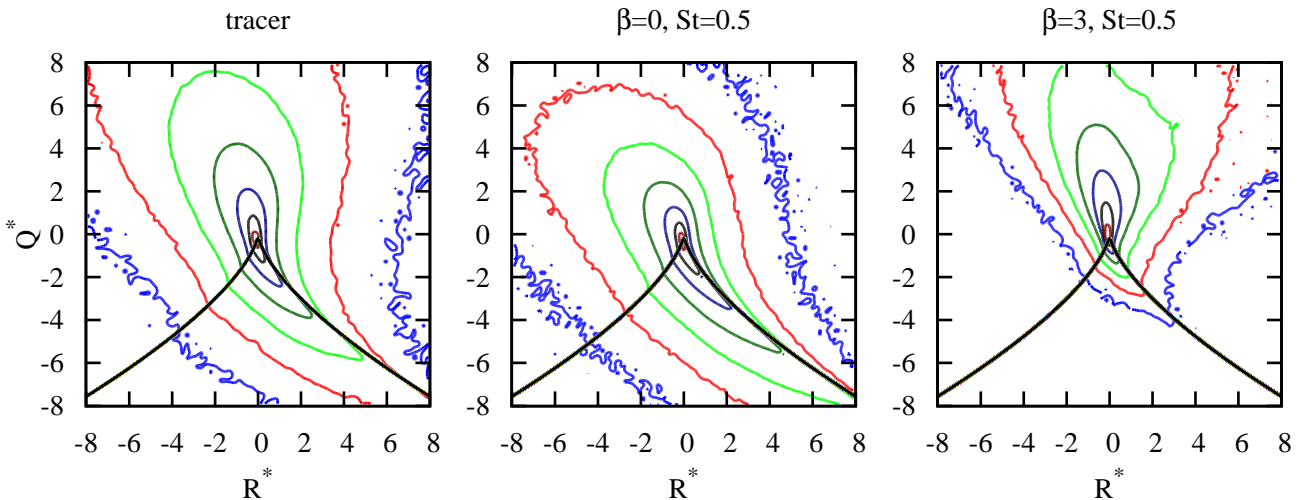}\vspace{-0.5cm} 
  \caption{Joint probability density function ${\mathcal P}(Q^*,R^*)$ of $Q^* \equiv Q/\langle Q^2 \rangle^{1/2}$ and $R^* \equiv R/ \langle Q^2 \rangle^{3/4}$, for particles of different types. Contour lines are drawn at values $10^{-z}$ with $z=0,1,2,3,4,5,6$ ( from the center to the outside of the figure). The thick line traces the curve: $ (R/2)^2 + (Q/3)^3= 0$ (Viellefosse  line), discriminating between complex (above) and real (below) eigenvalues of $\mathcal A$. Data come from $Re_{\lambda} = 180$ calculations.
  }
  \label{fig00}
\end{figure*}
\section{Refined Similarity Hypothesis in the Lagrangian Frame}\label{sec:1}
 
 \subsection{Inertia effect on the statistics of principal invariants of velocity gradient tensor} 
We have already noted that the effects of inertia may be of particular interest for
the present study. Inertial particles are not distributed
homogeneously in the volume, centrifugal force tends to concentrate
light particle inside strong elliptical regions, with high vorticity
\cite{eaton:1994,maxey:1987,calzavarini:2008a}; and heavy particle in
hyperbolic regions, typical of intense shear.
Following Chong et al. \cite{chong:1990}, the flow topology may be locally defined  in terms of the two principal invariants of the
velocity gradient tensor $\mathcal A =  A_{ij} = \partial_i u_j$, namely  $Q = - Tr[\mathcal A^2]/2$ and $R = - Tr[\mathcal A^3] / 3$, (see  also \cite{luethi:2009} for a recent study).
$Q$ represents the difference between a rotation-dominated and a dissipation-dominated flow topology, e.g., it is positive in a vortex core, while negative in a region characterized by high strain. The second parameter, $R$, analogously represents the  competition between the vorticity production and the dissipation production. Also, the separatrix curve $(R/2)^2+ (Q/3)^3=0$ (so called Vieillefosse line \cite{vieillefosse:1984}) discriminate between three real or one real and two complex-conjugate eigenvalues for $\mathcal A$, again meaning only-strain or vortical regions.
In fig. \ref{fig00} we show the joint probability density function $\mathcal P(Q,R)$ for different particle types as measured in the simulations  at $Re_{\lambda}= 180$. The most striking effect is for light particles ($\beta=3$), contrary to tracers and heavy particles ($\beta=0$)  they spend essentially all of their time in upper half-plane $Q>0$, meaning that they constantly trapped in vortical regions. 
 
\subsection{Time correlation of symmetric and antisymmetric component of velocity gradient tensor}  
\begin{figure}
  \includegraphics[width=\hsize]{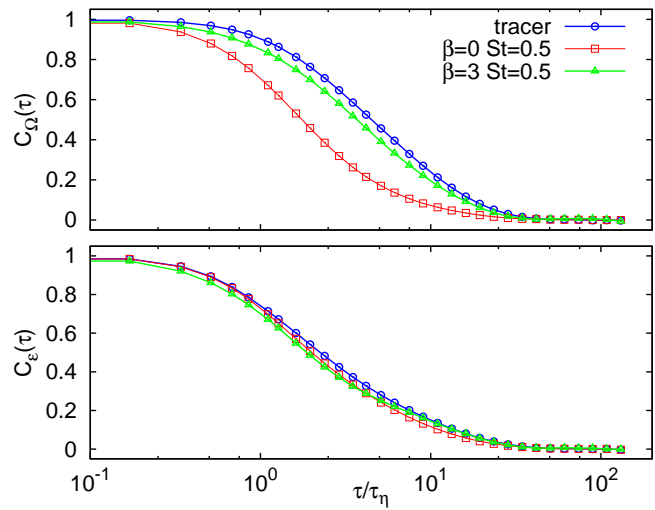}
  \caption{Temporal autocorrelation function, i.e., $C_X(\tau) \equiv
    \langle X'(t)-X'(t+\tau)\rangle/\langle X'^2\rangle$ with $X'(t) =
    X(t) - \langle X \rangle$, of the enstrophy $X=\Omega$ (top) and
    the energy-dissipation-rate $X=\varepsilon$ (bottom) for fluid tracers and for inertial particles with $St=0.5$, $\beta = 0 ,3$, at $Re_{\lambda} = 180$.}
  \label{fig-1}
\end{figure}

One also expects pretty different temporal correlations between particle trajectories and the
underlying topology of the carrier flow. We look now at the symmetric/antisymmetric component of the
velocity gradient tensor $\mathcal A$, because of their direct link with energy dissipation and enstrophy, \cite{yeung:2007,luthi:2005,guala:2007}.  
We show in Fig. (\ref{fig-1}) the autocorrelation function of enstrophy, $\Omega = \frac{\nu}{2} (\mathcal A - \mathcal A^T)^2 = \frac{\nu}{2}
\sum_{i,j}(\partial_i u_j -\partial_j u_i)^2$ and energy dissipation,
$\epsilon =  \frac{\nu}{2} (\mathcal A + \mathcal A^T)^2 =\frac{\nu}{2} \sum_{i,j}(\partial_i u_j+\partial_j u_i)^2$, along
the particle trajectories for different values of inertia. As one can
see both these quantities have short autocorrelation time, at $Re_{\lambda} = 180$ we find $T_{\Omega} = \int_{0}^{\infty} C_{\Omega}(\tau) d \tau   \simeq 7 \tau_{\eta}$ and similarly $T_{\varepsilon} \simeq 5 \tau_{\eta}$.
However, the autocorrelation of enstrophy turns out to be rather sensitive to the type of
particles, while energy dissipation is probed more or less uniformly.
This is a clear indication that due to inertia, particle tends to leave
in regions with very different vorticity contents, while energy dissipation - although different in intensity - turns out to be a more robust quantity  in term of coherence-in-time: This result will be very useful in our following discussion.

\subsection{RSH and its generalized formulation}
Along the trajectory of a fluid tracer $\bm x(t)$ the velocity difference will be
denoted as $\delta_\tau v = v(t+\tau) - v(t)$ and similarly we define a coarse grained
energy dissipation measured along the trajectory as
$\varepsilon_{\tau}= \int_t^{t+\tau}{\varepsilon(\bm x(t),t)\ dt}$ (see also \cite{borgas:1993}).
The RSH (\ref{eq:erksh}) can be translated from space to time by making the assumptions that $ \delta_{\tau} v \sim \delta_r u $ and 
$\varepsilon_{\tau} \sim \varepsilon_{r}$ when $\tau$ and $r$ are linked trough the eddy turnover time definition, $ \tau(r) \sim r/
\delta_r u$.
This argument leads to the Lagrangian refined similarity hypothesis (LRSH):
\begin{equation}
  \label{rksh_temp}
  \delta_\tau v \sim  \tau^{1/2}\, \varepsilon_{\tau}^{1/2}.
\end{equation}
In order to test Eq. (\ref{rksh_temp}) one should verify, for any exponent $p$, the scaling
relations:
\begin{equation}
  \langle (\delta_\tau v)^p \rangle \simeq \tau^{p/2} \langle \varepsilon_{\tau}^{p/2} \rangle,
\label{benzi1}
\end{equation}
where $\simeq$ means equal apart from a multiplicative constant depending only on $p$, in the inertial range.  In the time domain the inertial range is defined as the interval, $\tau_{\eta} \ll \tau \ll T_L$, where $T_L$ is the Lagrangian integral time scale, which is estimated as the autocorrelation time of velocity of fluid tracers, i.e., $T_L= \int_0^{\infty} C_{v}(\tau) \ d\tau$, with $C_{v}(\tau) \equiv  \langle  v(t)v(t+\tau) \rangle / \langle v^2\rangle$. As one can estimate $T_L/\tau_{\eta} \sim Re_{\lambda}$, the extension of the inertial range in dissipative time-scale units extends over roughly two decades in the present numerical study.

In contrast to the 4/5-law (consequence of the Karman-Howarth equation) leading to exact scaling properties for third order velocity increments in the Eulerian frame, we do not have any exact scaling relation derivable from NS equations in the Lagrangian domain.
Furthermore, it is known that in the Lagrangian frame, finite Reynolds
effects induce larger deviations from a power law regime than what
observed in Eulerian frame \cite{yeung:2002}.
To overcome these effects, following \cite{benzi:1996},
we can generalize the above expression (\ref{benzi1}) by using its Extended Self
Similarity (ESS) form, namely:
\begin{equation}
  \langle (\delta_\tau v)^p \rangle \simeq 
 \left( \frac{ \langle  (\delta_\tau v)^2 \rangle}
  {\langle \varepsilon_{\tau} \rangle} \right)^{p/2}  \langle \varepsilon_{\tau}^{p/2} \rangle
\label{benzi2}
\end{equation}

\subsection{Numerical tests of LRSH}
\begin{figure}
  \includegraphics[width=1.0\hsize]{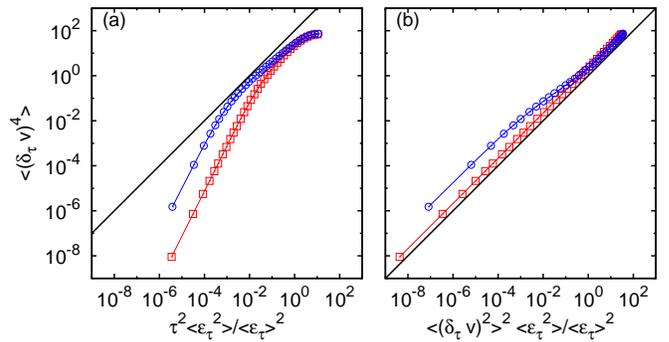}
  \caption{\label{fig0} Test of LRSH along the trajectories of
    tracers and heavy particles at $\mathrm{Re}_\lambda=400$.  (a) For
    $p=4$ we show Eq. (\ref{benzi1}) for $St=0$ (circles) and $St=2$
    (squares). (b) We show the validity of Eq. (\ref{benzi2}) for
    the same values of $p$ and $St$. Straight lines correspond to the
    theoretical scaling prediction. Data come from $Re_{\lambda} = 400$ calculations.}
\end{figure}

In Figure \ref{fig0}(a) we present a test of Eq.  (\ref{benzi1}) for
$p=4$ for particles with $\beta=1$, $St=0$, i.e. fluid tracers (circles) and
very-heavy particles with $\beta=0$, $St=2$ (triangles). In Figure \ref{fig0}(b), we show instead the relation from Eq.
(\ref{benzi2}) for the same particle types. Two major results emerge. 
First the LRSH, as expressed by Eq. (\ref{benzi2}) is well verified for the
transport of particles in turbulent flows. The use of the ESS version
for LRSH is able to overcome finite size/time effects which are usually
observed at relatively low Reynolds number (see
\cite{yeung:2002}).   The second important result, which will be
investigated later on in this manuscript, comes from inspecting the validity of
(\ref{benzi2}) for different Stokes numbers. Equation
(\ref{benzi2}) is supposed to be valid both in the inertial range and
in the dissipative range (where the velocity field is smooth) though with different offset. This is
clearly observed in Figure \ref{fig0}(b) for the case $St=0$.  
It is already known that by increasing the Stokes number,
particles tends to escape from strong vorticity region, thus
decreasing the effect of the dip present in between dissipative and
inertial scales \cite{bec:2006a}. As a consequence, for $St=2$ 
we observe almost no deviation of Eq. (\ref{benzi2}) in the range of scales between the inertial and
the dissipative ranges. 

\begin{figure}
  \includegraphics[width=\hsize]{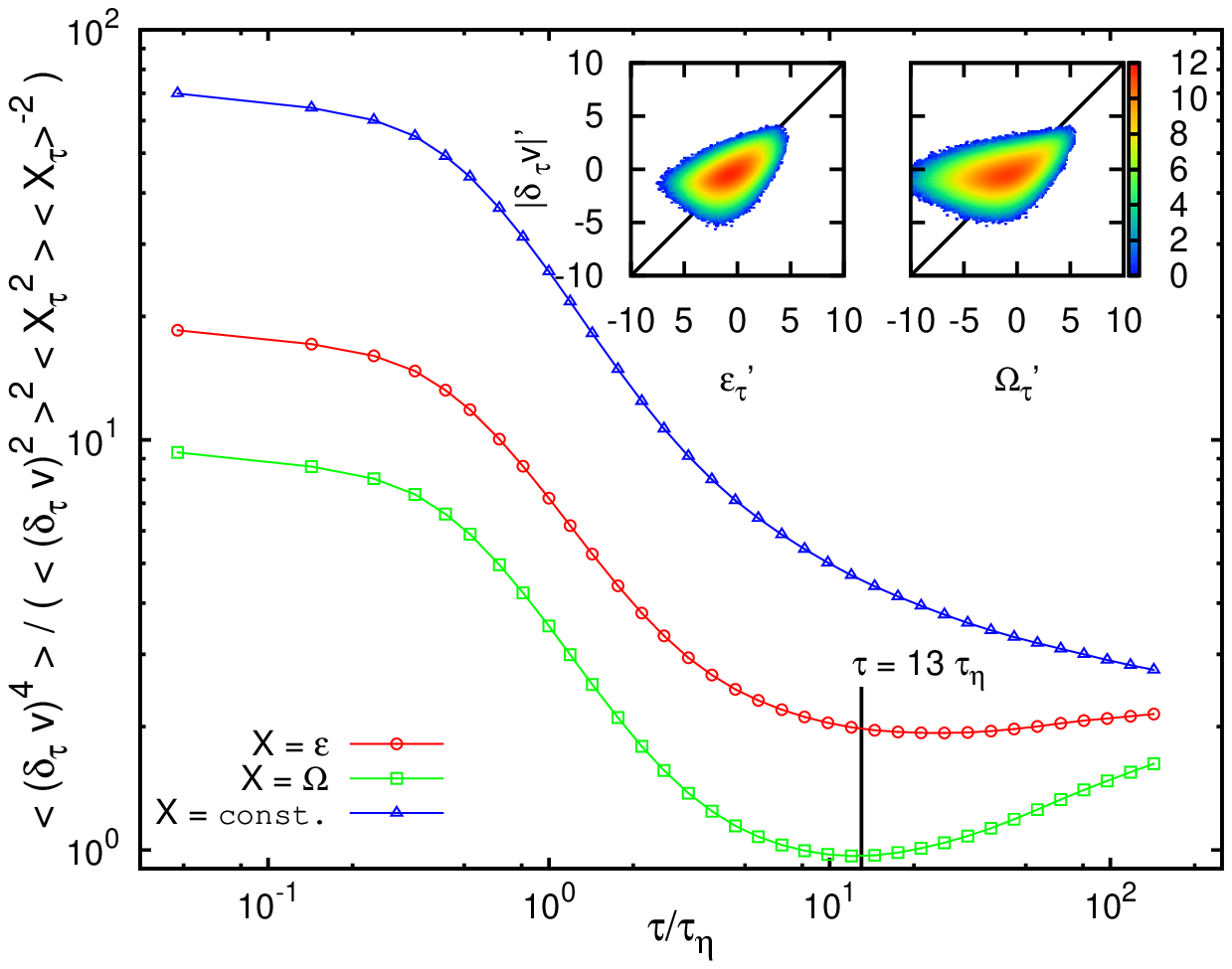}
  \caption{\label{fig1} Test of LRSH along the trajectories of
    tracers ($\mathrm{Re}_\lambda=400$). It is plotted $\left\langle (\delta_\tau v)^4\right\rangle / 
     ( \ \left\langle (\delta_\tau v)^2\right\rangle^2 {\left\langle X_\tau^{2}\right\rangle} {\left\langle X_\tau \right\rangle}^{-2}  )$:
     (circles) represent the case $X=\varepsilon$, 
     (squares) the case $X=\Omega$, and (triangles) the case $X=\mbox{const}$.
    In the inset the joint pdf:
     $P(\epsilon_\tau',| \delta _\tau v |'$) and $P(\Omega_\tau',| \delta_\tau v |'$) 
   at $\tau=13 \tau_{\eta}$; 
    (note that the prime symbol denotes variables normalized
    respect to their mean values, i.e. $x' \equiv x/\langle x \rangle$).}
\end{figure}

To have a more quantitative check,
we look now at the ratio between the two sides of Eq.\ (\ref{benzi2}),
namely at $\langle (\delta_\tau v)^p \rangle$ divided by  
$\langle \varepsilon_{\tau}^{p/2} \rangle \langle  (\delta_\tau v)^2 \rangle^{p/2} \langle \varepsilon_{\tau} \rangle^{-p/2}$, as a function of the time difference $\tau$. 
In Figure \ref{fig1} we show its behavior for the order $p=4$ fluid tracers particles,
the time difference $\tau$ is normalized by the dissipative time-scale $\tau_\eta$.
We observe  a plateau (see circles  symbols, in Fig \ref{fig1}) for $\tau/\tau_{\eta}\ge 5$. 
Notice that also in the dissipative range the compensation works well, as it should from the
requirement that the velocity field becomes differentiable, $\delta_{\tau} v \sim \tau$. However, the plotted ratio shows a mismatch
with the value attained in the inertial range. The transition between
the two plateaux occurs around the dissipative time scale, where the
presence of vortex trapping has been shown to  spoil the scaling behavior of
Lagrangian structure functions  $\langle (\delta_\tau v)^p
\rangle$ of the tracers  \cite{mazzitelli:2004,biferale:2005a,biferale:2008,arneodo:2008}.
In the same figure we show that using the coarse grained enstrophy,
i.e., $\Omega_\tau$ instead of $\varepsilon_\tau$, the compensation is
worse (squares). Similarly, compensation with
  enstrophy does not work neither for heavy nor for light
  particles (not shown).  Compensating without coarse grained
quantities, i.e. checking the deviation from dimensional,
non-intermittent, scaling does not provide a good plateau (triangles
in Fig \ref{fig1}).  This result supports the validity of LRSH only
when using the energy dissipation as the main driving process along
the particle motion. 
The behavior for intense fluctuations (moments higher then $p=6$) can not be checked quantitatively due
  to the lack of statistics. Nevertheless, in the same figure, we
show the joint probability density functions, ${\cal P}(\Omega_\tau,
|\delta_\tau v|)$ and ${\cal P}(\epsilon_\tau, |\delta_\tau v|)$, for
a time lag $\tau = 13\ \tau_\eta$. Velocity increments are more
correlated with coarse-grained energy than with enstrophy, as shown by
the high probability measured for simultaneous intense values of
$|\delta_\tau v|$ and $\varepsilon_\tau$.
\begin{figure}[t!]
  \includegraphics[width=\hsize]{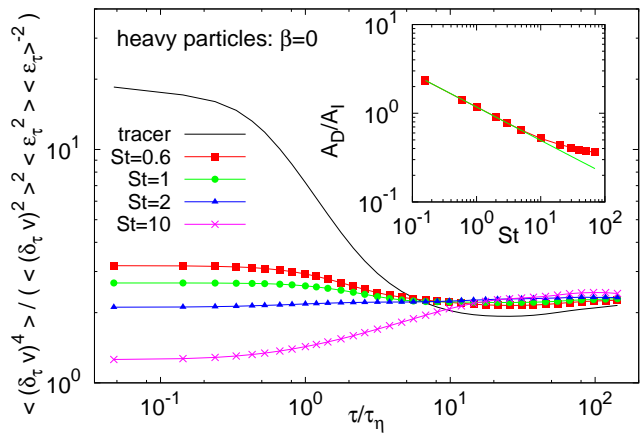}
  \caption{\label{fig2} Same as in Fig.\ref{fig1}, here also along the trajectories of
    heavy particles ($\mathrm{Re}_\lambda=400$). It is plotted 
     $\langle (\delta_\tau v)^4 \rangle \langle  \varepsilon_{\tau} \rangle^2  /
\left( \langle \varepsilon_{\tau}^2 \rangle \langle (\delta_\tau v)^2 \rangle^2 \right)$.
     Particles
    with $St=0.6, 1, 2, 10$ are compared with the result for
    tracers (solid black line). The LRSH is satisfied both in the
    inertial and in the dissipative ranges. The prefactors, $A_I$ and $A_D$, however differs in the two regions. 
    Notice that for  the largest Stokes, $St=10$, the smallest time lags where LRSH is verified, as expected, roughly $\tau \sim 10 \tau_\eta$. In the inset,
    the behavior of ratio of prefactors $A_D/A_I$ is plotted
    vs. $St$. For small $St$ values a behavior as $St^{-0.38}$ is
    found (solid line in the inset).}
\label{fig:2}
\end{figure}

Having established the validity of the LRSH, we strive now at
investigating the effect of different Stokes number and different density properties.

\section{LRSH in the ($\beta,St$) particle parameter space}\label{sec:2}

\subsection{Heavy particles ($\beta=0$) at  $St \sim O(1)$}
When particles have inertia their trajectories deviate from material
lines of the flow. In principle, one expects that for very small value
of the inertia (particles very close to fluid tracers) no appreciable
discrepancies can be measured.  
In Figure \ref{fig2} it is shown the test for
the LRSH compensated with the energy dissipation rate.  From
One can appreciate that in the inertial range,
e.g. $\tau/\tau_{\eta} \ge 5$ the LRSH is well verified for all the
Stokes considered. In the dissipative range it is also verified but
with a different proportionality constant. In particular, the
important mismatch observed between the two plateaux for tracers in
(Fig.\ref{fig1}) here is reduced considerable as soon as some inertia 
is switched on. This confirms that heavy particles are quickly
expelled out of vortex filaments, and therefore much less sensitive to
the transition around $\tau/\tau_\eta \sim 1$ than tracers
\cite{bec:2006a} (the opposite will happen for light particles, see
below).

The behavior of the ratio ($A_D/A_I$) of the plateaus displayed by
$\langle (\delta_\tau v)^4 \rangle \langle \varepsilon_{\tau}
\rangle^2 / \left( \langle \varepsilon_{\tau}^2 \rangle \langle
  (\delta_\tau v)^2 \rangle^2 \right)$ respectively in the dissipative
range ($A_D$ for $\tau \ll \tau_\eta$) and in the inertial range ($A_I$ for $\tau \leq 10 \tau_\eta$), is shown in the inset
of Fig.\ \ref{fig2}.  
  The estimate for the slope of $A_D/A_I$ vs.\ $St$ can be provided by the
  following reasoning. First we notice that the inertial constant
  $A_I$ is almost insensitive from the Stokes value, therefore the dissipative
  constant $A_D$ carries all the $St$ dependency. Moreover, we have
  measured that the single point energy dissipation statistics is pretty
  insensitive to the Stokes number (see again fig. \ref{fig-1}).
  As a consequence the main
  dependency on $St$ for the ratio $A_D/A_I$ comes from the flatness factor
  $F(\tau) \equiv \langle (\delta_\tau v)^4 \rangle / \langle (\delta_\tau
  v)^2\rangle^2$ in the intermediate-dissipative $\tau$ limit.  It is reasonable to
  estimate the difference between $F(\tau)$ at changing Stokes, but fixed Reynolds, as given by the value of the flatness
  at the particle response time: $F(\tau_p) \sim St^{\zeta_4-2\zeta_2}$,  where $\zeta_p$ is the
  $p$-th order scaling exponent for Lagrangian structure
  functions $\langle (\delta_\tau v)^4 \rangle \sim \tau^{\zeta_p}$. 
  Based on the experimental values $\zeta_4 \simeq 1.6$ and $\zeta_2 \simeq 1$ \cite{biferale:2008}, this estimate gives $A_D/A_I
  \sim St^{-0.4}$, not too far from the fit $St^{-0.38 \pm 0.05}$ to
  our numerical data, see inset of Fig \ref{fig:2}.  

  \subsection{Finite density contrast: heavy and light particles}
\begin{figure}
  \includegraphics[width=\hsize]{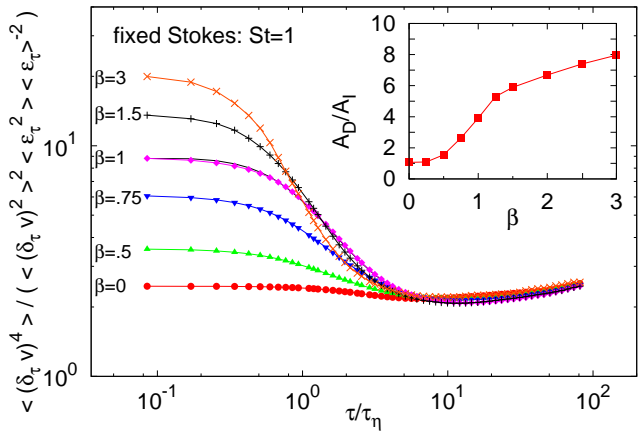}
  \caption{\label{fig3} Same as in fig. \ref{fig1} along the trajectories of
    heavy/light inertial particles with $St=1.0$ ($\mathrm{Re}_\lambda=180$). It is plotted
    $\langle (\delta_\tau v)^4 \rangle \langle  \varepsilon_{\tau} 
\rangle^2  /
\left( \langle \varepsilon_{\tau}^2 \rangle
 \langle (\delta_\tau v)^2 \rangle^2 \right)$.
  Particles with $\beta=0,0.5,0.75,1.0,1.5,3.0$ are
    compared with the result for tracers (solid black line). The LRSH
    is satisfied both in the inertial and in the dissipative
    ranges. As for the case of heavy particles the prefactors differs
    in the two regions. In the inset, the behavior of ratio of
    prefactors $A_D/A_I$ is plotted vs.\ $\beta$.}
\end{figure}

  We now look at the statistical properties of particles with finite
  density contrast, i.e. also $\beta \neq 0$.  In Figure \ref{fig3} it
  is shown, for $St=1$, the behavior of the compensated tests for 
  LRSH for different values of $\beta$ spanning the full range $[0 : 3]$.  
  In the inset it is also shown the behavior, this time as function of $\beta$, of the
  $A_D/A_I$ ratio. Notice how the critical value $\beta=1$,
  discriminating between heavy ($\beta <1$) and light ($\beta>1$)
  particles plays a crucial role. Again LRSH is well verified in the
  inertial range, but the change to a different plateau around
  $\tau/\tau_\eta \sim 1$ is now much more abrupt when light particles
  are considered: for those the vortex trapping is more pronounced, as
  all light particles quickly move towards high vorticity regions,
  showing a very sensitive dependency around the dissipative time
  dynamics. A model for the dependency of $A_D/A_I$ vs. $\beta$ is presently not available.

\begin{figure}[t!]
  \includegraphics[width=\hsize]{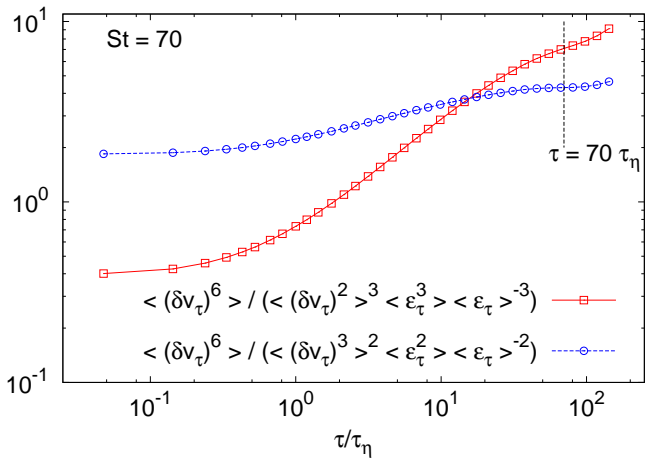}
  \caption{\label{fig4}Particles with very large inertia do not verify
    the LRSH (\ref{benzi2}), but follow the Eulerian version (\ref{eq:benzi3}).
    We show this for the order $p=6$ on particle trajectories with $St=70$, at $Re_{\lambda} = 400$.  
    As it can be seen, in the inertial range, the Eulerian RSH compensate better than LRSH.}
\end{figure}

\subsection{Heavy particles ($\beta=0$) with large inertia}
Having studied the case of particles with small inertia, we
now focus on the case of extreme inertia, i.e., when the response time of the
particle is at the top end of the inertial time-range, or $St \sim O(10)$. In this condition, for time lags $\tau < \tau_p$ 
when the particle filters out most of the underlying turbulent fluid fluctuations and evolves nearly ballistically,  one can predict a different behavior for $\delta_\tau v$. 
Along the trajectory of a ballistic particle, the relation linking scale to time is $\tau(r) \simeq r / v_0$ where the typical particle velocity $v_0$ is proportional to root mean square fluid velocity.  Recasting eq.(\ref{eq:erksh}) from space to time notation we obtain again an Eulerian-like RSH relation,
$(\delta_{\tau} v) \sim \left(\tau / v_0 \right)^{1/3}\ \varepsilon_{\tau}^{1/3}$, or:
\be 
\langle (\delta_{\tau} v)^p \rangle \simeq  \left( \frac{\tau}{v_0} \right)^{p/3} \langle \varepsilon_{\tau}^{p/3}  \rangle 
\label{eq:benzi3}
\ee
The generalized version of (\ref{eq:benzi3}) reads now
\be 
\langle (\delta_{\tau} v)^p \rangle \simeq  \frac{ \langle \varepsilon_{\tau}^{p/3} \rangle} {  \langle \epsilon_{\tau} \rangle^{p/3} }   \langle (\delta_{\tau} v)^3 \rangle^{p/3}
\label{eq:benzi4}
\ee
In Figure \ref{fig4} we present a test of this idea for $\langle (\delta_{\tau} v)^p \rangle$ with $p=6$ : For particles with very large Stokes numbers ($St=70$) we compensate the velocity increments both with respect to the prediction of the Lagrangian RSH and with respect to the
prediction of the Eulerian RSH in its generalized version. 
The compensation with the Eulerian RSH works appreciably better in the
range $\tau \leq St \cdot \tau_{\eta}$ than the compensation with LRSH.

\section{Conclusions}\label{sec:3}

In summary, some important statistical properties of velocity gradients along
trajectories of fluid tracers, heavy and light particles have been
investigated. We used high-resolution, high-statistic numerical data
to correlate the temporal properties of velocity gradients and velocity differences along
trajectories. We demonstrated that the Refined Similarity Hypothesis is well verified both for fluid particles and particles with response time in the dissipative regime, a feature that we dubbed Lagrangian RSH.  Around the dissipative time lags, heavy and light particles behave strongly
differently, due to the effect of being expelled/concentrated out/in vortex filaments. The dynamics at those time lags becomes markedly
dependent on the underlying topological flow properties.

Understanding the RSH in the Lagrangian domain may also have important applied consequences.  
In many applications, the geometry of the system and/or the intensity of turbulence do not allow for a direct
attack of the problem using numerical simulations of the Navier-Stokes equations. 
Modeling is needed for both the underlying fluid and for the particle equations. 
Typically, the ideal model, would like to replace Eqs.(\ref{eq:dynamics})-(\ref{eq:ns}) with a Langevin-like
equation for the particle evolution \cite{pope:2000,sawford:1991}:
${\rm d}\bm x /{\rm d} t=\bm v $, 
${\rm d}\bm v / {\rm d} t = {\mathcal D}(\mathcal A) {\bm v} + {\bm \Gamma}(t)$
where the $\Gamma$ represents some stochastic noise induced by the
underlying turbulent fluctuations. 
The hard physical problem is in the modelization of the drift term, ${\mathcal D}(\mathcal A)$, depending on the local
gradient structure along the trajectories (see
\cite{naso:2005,chevillard:2006a,biferale:2007a} for recent
attempts). Such term should also take into account effects induced by
preferential concentration in/out vortex filaments for the case of
inertial particles around the dissipative time lags.\\
The numerical database presented here can play a crucial role for benchmarking stochastic models for tracers and
inertial particles in turbulence. Data from this study are publicly available in unprocessed raw format from the iCFDdatabase (\url{http://cfd.cineca.it}).

During the preparation of this manuscript we got aware of a slightly similar investigation \cite{ym} where Lagrangian correlation of velocity and pressure gradients are studied conditioning on the initial  Eulerian energy dissipation, a sort of mixed Eulerian-Lagrangian refined Kolmogorov hypothesis, different from the fully Lagrangian view point adopted here.

We thank J. Bec, M. Cencini  and A. S. Lanotte  for several discussions. DEISA Consortium (co-funded by the EU, FP6 project 508830) is acknowledged for support within the DEISA Extreme Computing Initiative (\url{www.deisa.org}). LB acknowledges partial support
from CNISM. EC acknowledges support from CNRS and Agence Nationale de la Recherche.


\end{document}